%% file: arxiv.tex
\documentclass{article}
\usepackage{spconf,amsmath,graphicx,hyperref}
\usepackage{cite}
\usepackage{amssymb,amsfonts}
\usepackage{algorithmic}
\usepackage{textcomp}
\usepackage{xcolor}
\usepackage{multirow}
\usepackage{pifont}
\usepackage{makecell}
\usepackage{subcaption}
\usepackage{threeparttable}
\usepackage{booktabs}
\usepackage{url}
\usepackage[table]{xcolor}
\usepackage[ruled,vlined]{algorithm2e}

\newcommand{\cmark}{\ding{51}}
\newcommand{\xmark}{\ding{55}}
\title{Teaching LLMs to Hear Who Spoke What:\\Metadata-Supervised Pretraining for Encoder-Free Speech-LLMs}
\name{Mohan Shi$^{1,2*}$, Ruchao Fan$^2$, Sunit Sivasankaran$^2$, Keqi Deng$^2$, Jinyu Li$^2$\thanks{$^*$Work done during an internship at Microsoft.}}
\address{$^1$University of California, Los Angeles, USA \\ $^2$Microsoft CoreAI, Redmond, USA}
\begin{document}
\ninept
\maketitle
\begin{abstract}
Encoder-based speech large language models (Speech-LLMs) commonly employ pretrained speech encoders that prioritize linguistic content but may discard fine-grained acoustic cues essential for speaker discrimination and paralinguistic understanding. Encoder-free Speech-LLMs instead map Mel-spectrogram features directly into the LLM input space through lightweight embedding layers, enabling the LLM to learn from low-level acoustic features. However, systematic pretraining strategies for encoder-free Speech-LLMs remain underexplored, limiting their ability to compensate for the absence of large-scale pretrained speech encoders. We propose metadata-supervised pretraining (MSP), which leverages speech attributes such as speaker identity and emotion to develop speaker-discriminative and paralinguistic capabilities. We further introduce speaker-aware utterance composition (SAUC) to strengthen speaker discrimination and apply random span masking to regularize pretraining. We primarily evaluate our approach on joint ASR and speaker diarization in multi-speaker conversations, complemented by experiments on paralinguistic speech-understanding tasks. Under matched training-data conditions, our encoder-free model outperforms its randomly initialized encoder-based counterpart. With limited metadata-annotated data, it is competitive with models using speech encoders pretrained on substantially larger corpora, outperforming them in several settings. These results demonstrate the potential of encoder-free architectures for building native multimodal LLMs that acquire diverse speech capabilities.
\end{abstract}
\input{introduction}

\input{method}

\input{experiments}

\input{conclusion}

\newpage
\section{Generative AI Use Disclosure}
During the preparation of this work, the authors used large language models (Claude Opus 5) for language editing, including proofreading and improving clarity and readability of the manuscript. All technical content, experimental design, results, and conclusions were produced and verified by the authors. After the use of generative AI, the authors reviewed and edited the manuscript and take full responsibility for the content of the publication.


\end{document}

%% file: introduction.tex
\section{Introduction}
\label{sec:intro}

Speech large language models (Speech-LLMs)~\cite{FathullahWLJSLG24,ma2024embarrassingly,ShiJXXZWSZY24} typically consist of a pretrained speech encoder, a modality projector, and an LLM~\cite{grattafiori2024llama,yang2025qwen3}. The speech encoder, often based on Transformer~\cite{DongXX18} or Conformer~\cite{GulatiQCPZYHWZW20} architectures and pretrained on large-scale speech corpora, transforms speech into high-level representations that are mapped by the projector into the LLM embedding space. However, speech encoders in Speech-LLMs are often optimized with semantically oriented objectives~\cite{abouelenin2025phi,nvidia_canary_qwen_25b}, such as ASR, and therefore tend to prioritize linguistic content over fine-grained acoustic information. Consequently, their representations may not preserve sufficient acoustic cues for capturing diverse paralinguistic attributes, such as speaker identity and emotion. Previous studies have addressed this issue using dual-encoder architectures~\cite{TangYSC000M024,HuZ0CMHPL0SLW24,peng2026vibevoice} to separately capture semantic and acoustic information or auxiliary objectives~\cite{shi2026train} that encourage a single encoder to learn richer speech representations. Nevertheless, these approaches increase architectural or training complexity and typically require dedicated encoder training.

Recent works~\cite{team2026gemma,fan2026llm} have explored encoder-free architectures for Speech-LLMs, replacing heavy speech encoders with lightweight embedding layers that extract audio representations. This design not only simplifies the architecture but also allows the LLM itself to directly access low-level spectral features or raw audio, preserving the full acoustic information available in the input. The encoder-free model in Gemma~\cite{team2026gemma} demonstrates reasonable performance on ASR and speech translation, but does not evaluate paralinguistic capabilities. Mel-LLM~\cite{fan2026llm} further demonstrates promising performance on ASR and paralinguistic audio-understanding tasks without relying on a speech encoder pretrained on large-scale data. Nevertheless, it does not introduce an explicit training strategy to promote the acquisition of fine-grained speaker-discriminative or paralinguistic representations. Moreover, performance on audio-understanding benchmarks may also reflect instruction-following and reasoning abilities, making it difficult to isolate the model's ability to capture such information. It therefore remains unclear to what extent encoder-free Speech-LLMs can acquire and retain speaker-related and other paralinguistic information from low-level acoustic features.

To address this gap, we introduce metadata-supervised pretraining (MSP) for encoder-free Speech-LLMs, using speech attributes such as speaker identity and emotion to promote speaker discrimination and paralinguistic understanding. 
We further introduce speaker-aware utterance composition (SAUC) to strengthen speaker discrimination by constructing multi-speaker sequences with simulated overlaps and silence gaps, and apply random span masking to regularize pretraining. After pretraining, we separately fine-tune the model for each downstream task. We primarily evaluate our approach on joint ASR and speaker diarization~\cite{shi2026train}, which requires transcribing speech and attributing it to the correct speakers in multi-speaker conversations. We use a fixed prompt across all examples to reduce instruction-following effects. As complementary evaluation, we consider several paralinguistic speech understanding tasks following~\cite{fan2026llm}, including emotion, age, and gender recognition, as well as speaker verification. Under matched training-data conditions, our encoder-free model outperforms its randomly initialized encoder-based counterpart overall. Moreover, with limited metadata-supervised data, it is competitive with encoder-based models whose speech encoders are pretrained on substantially larger datasets, outperforming them in several settings.

To our knowledge, this work makes the following contributions:
\begin{itemize}
    \item We conduct the first systematic investigation of metadata-supervised pretraining for encoder-free Speech-LLMs, demonstrating its effectiveness in developing speaker-discriminative and paralinguistic capabilities.

    \item We introduce speaker-aware utterance composition (SAUC) to improve speaker discrimination in multi-speaker scenarios and random span masking to regularize metadata-supervised pretraining.

    \item We are the first to demonstrate that Speech-LLMs without pretrained speech encoders can achieve competitive performance on joint ASR and speaker diarization as well as paralinguistic understanding tasks, outperforming encoder-based counterparts under matched training-data conditions.
\end{itemize}

%% file: method.tex
\section{Method}
\label{sec:method}

\begin{figure}[t]
  \centering
  \includegraphics[width=\linewidth]{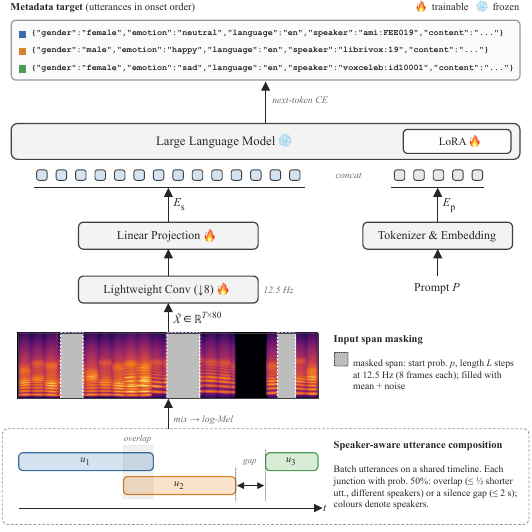}
  \vspace{-0.5cm}
  \caption{Proposed Metadata-supervised pretraining (MSP) of the encoder-free Speech-LLM.}
  \label{fig:msp}
  \vspace{0.3cm}
\end{figure}

We retain the encoder-free backbone of Mel-LLM~\cite{fan2026llm} and focus on its pretraining strategy, illustrated in Fig.~\ref{fig:msp}. We first perform metadata-supervised pretraining (MSP), then fine-tune the pretrained model separately for joint ASR and speaker diarization and for paralinguistic speech understanding using the corresponding task-specific targets.

\subsection{Encoder-free Speech-LLM}
\label{ssec:backbone}

Given an utterance waveform, we extract an $80$-dimensional log-Mel spectrogram $X\in\mathbb{R}^{T\times 80}$, and downsample it by a factor $r$ with lightweight convolution layers. A linear projection maps the result into the LLM embedding space,
\begin{equation}
E_{s}=\mathrm{Proj}\!\left(\mathrm{Conv}_{r}\!\right)
\in\mathbb{R}^{N\times d_{\mathrm{LLM}}},\quad N=\lceil T/r\rceil ,
\label{eq:frontend}
\end{equation}
with $r=8$, i.e.\ a $12.5$\,Hz token rate. No speech encoder blocks are used. The speech embeddings are concatenated with the prompt embeddings $E_{p}$ and consumed by the LLM, which is frozen and adapted with LoRA~\cite{HuSWALWWC22}. All pre-training and fine-tuning stages in this work share this frontend, so every comparison we report differs only in the training objective or the training data.

\subsection{Metadata-supervised Pretraining}
\label{ssec:meta}

Removing the encoder exposes the LLM directly to the Mel spectrogram, which retains low-level acoustic detail that a semantic encoder tends to compress away~\cite{fan2026llm}. Transcription supervision alone does not explicitly encourage learning speaker identity, emotion, or other paralinguistic attributes. We therefore pretrain the encoder-free model to predict these attributes alongside transcription, providing direct supervision for learning beyond lexical content.

\subsubsection{Meta labels}
\label{sssec:meta}
For each utterance $u_i$, we define a metadata record $m_i$ containing gender, emotion, language, speaker identity, and content (transcription). We serialize this record as a single JSON line with the fields in a fixed order,
\begin{equation}
\begin{split}
\pi=(\texttt{gender},\ \texttt{emotion},\
       &\texttt{language},\ \texttt{speaker},\ \texttt{content})
\end{split}
\label{eq:order}
\end{equation}
so that the paralinguistic attributes precede the lexical content. The \texttt{speaker} field is a global identifier, namespaced by source corpus so that it is unique across the entire training set, e.g.\ \texttt{ami:FEE019}.

\subsubsection{Speaker-aware Utterance Composition}
\label{sssec:stream}
Utterance-level corpora provide only short, single-speaker inputs, limiting the learning of speaker discrimination during pretraining. We therefore propose speaker-aware utterance composition (SAUC; Fig.~\ref{fig:msp}, bottom). We construct each training example on the fly from utterances within a training batch by concatenating them sequentially on a shared timeline and mixing them into a single audio stream, while composing their metadata targets in the corresponding temporal order. This synthesizes multi-speaker, long-form conditions from utterance-level data and makes the task more challenging: the model must distinguish multiple speakers within a single input rather than describe a single voice.

Strictly concatenating utterances end to end would make segmentation trivial, since every boundary would coincide with an acoustic discontinuity. Instead, the placement at each junction is sampled independently. With probability $0.5$, the next utterance starts before the current one ends, with an overlap of up to half the duration of the shorter utterance; otherwise, a silence gap of up to $2$,s is inserted. Overlaps are allowed only between different speakers, and the sequential construction ensures that at most two speakers are active at any time. Log-Mel features are extracted from the resulting stream, and the meta target lists the utterances in order of onset.

SAUC reuses utterance-level metadata to construct the corresponding targets automatically, avoiding the need to prepare and annotate a separate corpus of multi-speaker mixtures in advance.

\subsubsection{Input Masking}
\label{sssec:mask}

To regularize metadata-supervised pretraining, we apply span masking to the input features during pretraining (grey spans in Fig.~\ref{fig:msp}). Masking is defined at the $12.5$\,Hz temporal resolution of the convolutional frontend and then mapped to the corresponding log-Mel frames. Each position independently starts a masked span with probability $p$, and each span covers $L$ consecutive positions, with overlapping spans merged. Each masked position corresponds to eight consecutive log-Mel frames. These frames are replaced by the per-utterance, per-channel mean plus Gaussian noise scaled by a small fraction of the corresponding channel-wise standard deviation. Masking is applied only during pretraining.




%% file: experiments.tex
\section{Experimental Settings}
\subsection{Metadata-Supervised Pretraining}

\subsubsection{Pretraining Data}
The metadata corpus comprises the training sets of AMI (IHM)~\cite{CarlettaABFGHKKKKLLLMPRW05}, ICSI~\cite{JaninBEEGMPPSSW03}, TED-LIUM~\cite{tedlium3}, Earnings-22~\cite{earnings22}, VoxPopuli~\cite{voxpopuli}, Common Voice~\cite{commonvoice}, LibriSpeech~\cite{PanayotovCPK15}, VoxCeleb~\cite{NagraniCXZ20,ChungNZ18}, and Fisher~\cite{cieri2004fisher}, together with the medium subset of LibriHeavy~\cite{KangYYKYGLP24}, totaling approximately 10k hours. All these corpora provide speaker identity annotations. For utterances without emotion or gender annotations, we generate pseudo-labels using emotion2vec\_plus\_large\footnote{\url{https://huggingface.co/emotion2vec/emotion2vec_plus_large}\label{emo2vec}} and an ECAPA-based gender classifier\footnote{\url{https://huggingface.co/griko/gender_cls_svm_ecapa_voxceleb}\label{gender}}, respectively. Since VoxCeleb does not provide transcriptions, we generate pseudo-transcriptions using Whisper large-v3 through the faster-whisper implementation.\footnote{\url{https://github.com/SYSTRAN/faster-whisper}\label{whisper}}
During pretraining, we mix metadata examples with transcription-only examples from SPGISpeech~\cite{spgispeech}, MLS-English~\cite{mls}, GigaSpeech~\cite{gigaspeech}, and FLEURS-English~\cite{fleurs} at a $3{:}1$ ratio to preserve general ASR capabilities. Metadata and ASR examples use their respective task-specific prompts. Hereafter, we refer to this mixed-task training scheme as metadata-supervised pretraining (MSP).

We also investigate scaling pretraining through an additional ASR initialization stage, consisting of one epoch on one million hours of internal data, followed by MSP. We further expand the metadata corpus with the English subsets of Emilia and Emilia-YODAS~\cite{he2025emilia}, collectively denoted as Emilia-en.

\subsubsection{Pretraining Setup}

Our model is built upon Phi-4-MM~\cite{abouelenin2025phi}, using the encoder-free architecture described in Section~\ref{ssec:backbone}. Throughout pretraining and downstream fine-tuning, only the convolutional frontend, modality projector, and LoRA parameters are updated. Batches use a per-GPU budget of $8{,}192$ speech and text tokens, with $80$\,ms of speech per token. Metadata sequences contain roughly $40$ utterances composed using SAUC (Section~\ref{sssec:stream}) within this budget; interleaved ASR examples are not concatenated. We pretrain for 80k steps on 16 H200 GPUs using AdamW with a learning rate of $1\times10^{-4}$, $\beta=(0.9,0.95)$, $\epsilon=1\times10^{-7}$, and weight decay of $0.01$. The learning rate is linearly warmed up for 500 steps and then linearly decayed. Input masking uses $p=0.04$ and $L=4$ (320\,ms), corresponding to an expected masked fraction of $1-(1-p)^L\approx15\%$.

\subsection{Downstream Fine-tuning and Evaluation}
\subsubsection{Joint ASR and Speaker Diarization}


Following~\cite{shi2026train}, we fine-tune on approximately 10k hours of multi-speaker speech segmented into 15--20\,s clips, using the same target format.
Our encoder-free fine-tuning uses only the LLM prediction loss, without requiring the encoder-side auxiliary speaker-decoder loss of~\cite{shi2026train}.
We evaluate on 10--20\,s clips from the AMI~\cite{CarlettaABFGHKKKKLLLMPRW05} (IHM-Mix) test set and CH109 Full~\cite{LDC97S42}, following~\cite{shi2026train}. We use a fixed prompt across all examples, which reduces instruction-following effects. We report word diarization error rate (WDER)~\cite{ShafeySS19} for diarization and concatenated minimum-permutation word error rate (cpWER)~\cite{watanabe2020chime} for joint diarization and ASR, both expressed as percentages.

\subsubsection{Paralinguistic Speech Understanding}
\vspace{-0.1cm}

We fine-tune on approximately 10k hours of speech- and audio-understanding data, including audio and music instruction data, VoxCeleb speaker data, IEMOCAP~\cite{iemocap} (Sessions~1--4) and MOSEI emotion data, AudioSet/FSD50K/AudioCaps/Clotho-style data from LTU-AS~\cite{ltuas}, and reproduced WavLLM~\cite{HuZ0CMHPL0SLW24} speech and audio question-answering data. Following~\cite{fan2026llm}, we evaluate on IEMOCAP emotion recognition (Session~5), VoxCeleb gender and age prediction, and WavLLM speaker verification, excluding evaluation samples from pretraining and finetuning. We report accuracy for all tasks except age prediction, which uses mean absolute error (MAE).

\vspace{-0.25cm}
\subsection{Baselines}
\vspace{-0.15cm}
We consider three groups of baselines: (1) an encoder-based model with a randomly initialized speech encoder, trained using the same pretraining setup as our model; (2) an encoder-free model with the same architecture as ours but without metadata-supervised pretraining; and (3) previously reported results from related works~\cite{shi2026train,fan2026llm}.
\begin{table}[t]
\centering
\caption{Joint ASR and speaker diarization results on AMI and CH109. ``Audio tower'' denotes the speech encoder or convolutional frontend. MSP denotes metadata-supervised pretraining. Only the JEDIS-LLM models use the encoder-side auxiliary speaker-decoder loss; the loss ablation excludes it. The reproduced JEDIS-LLM results use text normalization during metric computation, which accounts for differences from the reported results.}
\vspace{-0.3cm}
\label{tab:main_results}
\renewcommand{\arraystretch}{1.15}
\resizebox{\columnwidth}{!}{
\begin{tabular}{lcccccc}
\toprule
\multirow{2}{*}{Model} &
\multirow{2}{*}{\makecell{Audio\\tower}} &
\multirow{2}{*}{MSP} &
\multicolumn{2}{c}{AMI} &
\multicolumn{2}{c}{CH109} \\
\cmidrule(lr){4-5}
\cmidrule(lr){6-7}
& & & WDER & cpWER & WDER & cpWER \\
\midrule

\multicolumn{7}{l}{\textit{Prior work}} \\
JEDIS-LLM~\cite{shi2026train} (reported)
    & \multirow{3}{*}{\makecell{Pretrained\\encoder}} & \xmark
    & 6.97 & 23.13 & \textbf{2.06} & 19.46 \\
JEDIS-LLM (reproduced)
    & & \xmark
    & 6.00 & 21.04 & 2.46 & 16.73 \\
\quad - speaker-decoder loss &  & \xmark & 10.43 & 25.59 & 3.62 & 17.76 \\

\midrule
\multicolumn{7}{l}{\textit{Controlled comparison}} \\
Encoder-based
    & \multirow{2}{*}{\makecell{Random-init\\encoder}} & \xmark
    & 13.48 & 43.02 & 6.28 & 36.76 \\
Encoder-based
    & & \cmark
    & 5.79 & 21.79 & 2.85 & 19.86 \\

\addlinespace[3pt]

Encoder-free
    & \multirow{5}{*}{\makecell{Lightweight\\convolution}} & \xmark
    & 6.94 & 25.14 & 3.34 & 21.90 \\
\textbf{Encoder-free (ours)}
    & & \cmark
    & 5.38 & 21.27
    & 2.77 & 19.08 \\
\quad + mask
    & & \cmark
    & \textbf{5.27} & 20.98
    & 2.65 & 18.92 \\
\quad \quad + 1M hrs ASR & & \cmark & 5.62 & 20.28 & 2.68 & \textbf{17.79} \\
\quad \quad \quad + Emilia-en MSP & & \cmark & 5.57 & \textbf{19.56} & 2.48 & 18.24 \\
\bottomrule
\end{tabular}}
\end{table}

\vspace{-0.4cm}
\section{Experimental Results}
\vspace{-0.2cm}
\subsection{Joint ASR and Speaker Diarization}
\vspace{-0.1cm}
Table~\ref{tab:main_results} reports joint ASR and diarization results. Under matched training-data conditions, the encoder-free model outperforms its randomly initialized encoder-based counterpart on all four metrics, both with and without MSP, despite eliminating the speech encoder. MSP substantially improves both architectures, supporting the effectiveness of metadata supervision for speaker-attributed transcription.


Without the auxiliary speaker-decoder loss, JEDIS-LLM yields higher WDER than the encoder-free baseline without MSP on both AMI ($10.43$ vs.\ $6.94$) and CH109 ($3.62$ vs.\ $3.34$), despite using a pretrained speech encoder. This result is consistent with our concern that semantically oriented encoders may attenuate acoustic cues important for speaker discrimination, highlighting the potential benefit of allowing the LLM itself to interpret low-level spectral features.

Input masking further improves all metrics, allowing encoder-free MSP to outperform reproduced JEDIS-LLM on both AMI metrics. Adding the 1M-hour ASR initialization lowers cpWER on both sets but slightly increases WDER, suggesting that large-scale transcription supervision does not necessarily improve speaker attribution. Scaling MSP with Emilia-en achieves the lowest AMI cpWER and a CH109 WDER comparable to reproduced JEDIS-LLM.

\begin{table}[t]
\centering
\caption{Ablation of pretraining objectives and SAUC. Pure ASR and MSP use the same training utterances. ``Overlap \& gap'' denotes simulated speech overlaps and silence gaps. Input masking is disabled for all variants.}
\vspace{-0.2cm}
\label{tab:sauc_ablation}
\renewcommand{\arraystretch}{1.12}
\resizebox{\columnwidth}{!}{
\begin{tabular}{ccccccc}
\toprule
\multirow{2}{*}{\makecell{Pretraining\\objective}} &
\multirow{2}{*}{\makecell{Utterance\\concatenation}} &
\multirow{2}{*}{\makecell{Overlap\\\& gap}} &
\multicolumn{2}{c}{AMI} &
\multicolumn{2}{c}{CH109} \\
\cmidrule(lr){4-5}\cmidrule(lr){6-7}
& & & WDER & cpWER & WDER & cpWER \\
\midrule
None
    & -- & --
    & 6.94 & 25.14 & 3.34 & 21.90 \\
pure ASR
    & \xmark & \xmark
    & 6.94 & 23.47 & 3.10 & 20.67 \\
\addlinespace[2pt]
MSP
    & \xmark & \xmark
    & 6.09 & 22.48 & 3.01 & 19.87 \\
MSP
    & \cmark & \xmark
    & 6.05 & 22.40 & 3.18 & 20.59 \\
MSP
    & \cmark & \cmark
    & \textbf{5.38} & \textbf{21.27}
    & \textbf{2.77} & \textbf{19.08} \\
\bottomrule
\end{tabular}}
\vspace{0.2cm}
\end{table}

\begin{table}[t]
\centering
\caption{Effect of metadata target ordering and timestamp supervision. Timestamp targets follow the format \texttt{start\textasciitilde end}. G, E, L, S, C, and T denote gender, emotion, language, speaker identity, transcription content, and timestamps, respectively. All variants are pretrained without input masking.}
\vspace{-0.2cm}
\label{tab:metadata_order}
\resizebox{0.85\columnwidth}{!}{
\begin{tabular}{lcccc}
\toprule
\multirow{2}{*}{Metadata target sequence} &
\multicolumn{2}{c}{AMI} &
\multicolumn{2}{c}{CH109} \\
\cmidrule(lr){2-3}\cmidrule(lr){4-5}
& WDER & cpWER & WDER & cpWER \\
\midrule
G--E--L--C--S & 5.61 & 21.88 & 2.94 & 20.08 \\
T--G--E--L--S--C
    & 5.40 & 21.76 & \textbf{2.77} & 19.57 \\
G--E--L--S--C (default)
    & \textbf{5.38} & \textbf{21.27}
    & \textbf{2.77} & \textbf{19.08} \\
\bottomrule
\end{tabular}}
\vspace{0.1cm}
\end{table}

\subsection{Ablation of Pretraining Objectives and SAUC}

Table~\ref{tab:sauc_ablation} examines the effects of the pretraining objective and SAUC. Pure ASR and all MSP variants use the same underlying training utterances, differing only in supervision and composition. Pure ASR pretraining improves cpWER on both sets but leaves AMI WDER unchanged. Utterance-level MSP improves all four metrics over pure ASR, demonstrating the value of metadata supervision beyond transcription. Concatenation alone yields marginal gains on AMI and degrades both CH109 metrics. Adding overlaps and silence gaps improves all four metrics, reducing WDER from $6.05$ to $5.38$ on AMI and from $3.18$ to $2.77$ on CH109. This suggests that varied speaker transitions and overlaps help the model learn speaker discrimination under conversational conditions. These gains require only on-the-fly composition, without a prepared corpus of multi-speaker mixtures.

\vspace{-0.1cm}
\subsection{Metadata Target Design}

Table~\ref{tab:metadata_order} compares the default target sequence in Eq.~\ref{eq:order} with alternative field ordering and timestamp supervision. Moving transcription before speaker identity degrades all four metrics. Preceding text may expose speaker-correlated lexical cues, encouraging shortcuts that weaken the learning of acoustic speaker representations. Adding timestamps also increases cpWER on both sets without improving WDER. Predicting precise decimal-valued timestamps in the \texttt{start\textasciitilde end} format may impose an additional learning burden, diverting capacity from other metadata attributes and transcription.

\vspace{-0.1cm}
\subsection{Input Span Masking}


Table~\ref{tab:mask_probability} compares masking probabilities with a fixed span length of $L=4$. Moderate masking ($p=0.04$, approximately $15\%$ of the input) achieves the lowest WDER and cpWER on both sets. Lighter masking ($p=0.01$) improves three metrics but slightly increases CH109 cpWER, while aggressive masking ($p=0.15$, approximately $48\%$) degrades both AMI metrics and CH109 cpWER. Thus, moderate masking encourages learning from partially observed speech, whereas excessive masking may remove acoustic evidence needed for transcription and speaker discrimination.

\begin{table}[t]
\centering
\caption{Effect of the span-masking probability $p$ with span length $L=4$ (320 ms). The expected overall masking ratio is $1-(1-p)^L$.}
\vspace{-0.2cm}
\label{tab:mask_probability}
\renewcommand{\arraystretch}{1.1}
\resizebox{0.85\columnwidth}{!}{
\begin{tabular}{lccccc}
\toprule
\multirow{2}{*}{\makecell{Masking\\probability $p$}} &
\multirow{2}{*}{\makecell{Masked\\ratio}} &
\multicolumn{2}{c}{AMI} &
\multicolumn{2}{c}{CH109} \\
\cmidrule(lr){3-4}\cmidrule(lr){5-6}
& & WDER & cpWER & WDER & cpWER \\
\midrule
0
    & 0\%
    & 5.38 & 21.27 & 2.77 & 19.08 \\
0.01
    & $\sim$4\%
    & 5.36 & 20.99
    & 2.68 & 19.15 \\
0.04 (default)
    & $\sim$15\%
    & \textbf{5.27} & \textbf{20.98}
    & \textbf{2.65} & \textbf{18.92} \\
0.15
    & $\sim$48\%
    & 5.63 & 21.56 & 2.76 & 19.78 \\
\bottomrule
\end{tabular}}
\vspace{0.3cm}
\end{table}

\begin{table}[t]
\centering
\caption{Paralinguistic speech understanding results. ACC denotes accuracy (\%) and MAE denotes mean absolute error. MSP uses the default input masking configuration.}
\vspace{-0.2cm}
\label{tab:paralinguistic_results}
\renewcommand{\arraystretch}{1.12}
\resizebox{\columnwidth}{!}{
\begin{tabular}{lcccc}
\toprule
Model &
\makecell{Emotion\\ACC $\uparrow$} &
\makecell{Gender\\ACC $\uparrow$} &
\makecell{Age\\MAE $\downarrow$} &
\makecell{Speaker\\ACC $\uparrow$} \\
\midrule
\multicolumn{5}{l}{\textit{Prior work}} \\
Mel-LLM~\cite{fan2026llm} (w/o ASR init.)
    & 66.16 & 97.73 & 8.39 & 87.98 \\
Mel-LLM (w/ ASR init.)
    & 72.36 & 98.21 & 7.56 & 92.10 \\
\midrule
\multicolumn{5}{l}{\textit{Controlled comparison}} \\
Random encoder
    & 51.65 & 97.38 & 8.25 & 74.41 \\
Random encoder + MSP
    & 72.28 & 98.12 & 7.34 & 91.07 \\
Encoder-free (Mel-LLM reprod., w/o ASR init.)
    & 64.30 & 97.07 & 8.42 & 85.81 \\
Encoder-free + MSP
    & 75.66 & 98.12 & 7.57 & 93.16  \\
\quad + 1M hrs ASR & 76.55 & \textbf{98.43} & \textbf{6.83} & \textbf{95.18} \\
\quad \quad + Emilia-en MSP & \textbf{77.92} & 98.25 & 6.95 & 94.99 \\ 
\bottomrule
\end{tabular}}
\vspace{0.2cm}
\end{table}

\subsection{Paralinguistic Speech Understanding}
\vspace{-0.1cm}
Table~\ref{tab:paralinguistic_results} reports results on paralinguistic speech understanding. MSP improves all four tasks over the reproduced encoder-free baseline, particularly emotion recognition and speaker verification. Under matched training-data conditions, encoder-free MSP outperforms encoder-based MSP on these two tasks, matches its gender accuracy, and yields slightly higher age MAE. These results support the effectiveness of metadata supervision within the simpler encoder-free architecture.
Large-scale ASR initialization further improves all four tasks, while adding Emilia-en to MSP yields the best emotion-recognition accuracy but slightly worsens the other three metrics. The consistent gains from ASR initialization suggest that these tasks may partly rely on lexical and semantic cues recoverable through ASR, rather than exclusively on acoustic or paralinguistic information. Improved acoustic representations from ASR pretraining may also contribute to these gains.

%% file: conclusion.tex
\vspace{-0.1cm}
\section{Conclusion}
\label{sec:conclusion}
\vspace{-0.1cm}
In this work, we presented metadata-supervised pretraining (MSP) for encoder-free Speech-LLMs. Rich metadata targets promote speaker discrimination and paralinguistic understanding beyond transcription. Speaker-aware utterance composition (SAUC) combines utterances into long-form multi-speaker inputs with random overlaps and silence gaps, simulating conversational conditions to strengthen speaker discrimination. Moderate input masking further regularizes MSP and improves downstream performance. Experiments on joint ASR and speaker diarization, together with paralinguistic understanding tasks, demonstrate the effectiveness of this approach. Under matched training-data conditions, the encoder-free model outperforms its randomly initialized encoder-based counterpart overall, with a simpler architecture and fewer parameters due to the removal of the speech encoder. Scaling ASR and metadata pretraining provides further, task-dependent gains. These results demonstrate the potential of encoder-free architectures for building native multimodal LLMs that acquire diverse speech capabilities through metadata-supervised pretraining.